\documentclass[epj]{svjour}
\usepackage{latexsym}
\usepackage{graphicx}   
\usepackage{epstopdf}   
\begin{document}
\title{On a nonlinear electromechanical model of nerve}
\author{Alain M. Dikand\'e\thanks{Email address: dikande.alain@ubuea.cm}
}                     

%
\institute{Laboratory of Research on Advanced Materials and Nonlinear Sciences (LaRAMaNS), Department of Physics, Faculty of Science, University of Buea, P.O. Box 63 Buea, Cameroon}
\date{Received: date / Revised version: date}
%
\abstract{
The generation of action potential brings into play specific mechanosensory stimuli manifest in the variation of membrane capacitance, resulting from the selective membrane permeability to ions exchanges and testifying to the central role of electromechanical processes in the buildup mechanism of nerve impulse. As well established [See e.g. D. Gross et al, Cellular and Molecular Neurobiology vol. 3, p. 89 (1983)], in these electromechanical processes the net instantaneous charge stored in the membrane is regulated by the rate of change of the net fluid density through the membrane, orresponding to the difference in densities of extacellular and intracellular fluids. An electromechanical model is proposed for which mechanical forces are assumed to result from the flow of ionic liquids through the nerve membrane, generating pressure waves stimulating the membrane and hence controlling the net charge stored in the membrane capacitor. The model features coupled nonlinear partial differential equations: the familiar Hodgkin-Huxley's cable equation
for the transmembrane voltage in which the membrane capacitor is now a capacitive diode, and the Heimburg-Jackson's nonlinear hydrodynamic equation for the pressure wave controlling the total charge in the membrane capacitor. In the stationary regime, the Hodgkin-Huxley cable equation with variable capacitance reduces to a linear operator problem with zero eigenvalue, the bound states of which can be obtained exactly for specific values of characteristic parameters of the model. In the dynamical regime, numerical simulations of the modified Hodgkin-Huxley equation lead to a variety of typical figures for the transmembrane voltage, reminiscent of action potentials observed in real physiological contexts.   
%
} 
\maketitle
\section{Introduction}
\label{intro}
\smallskip
The generation of nerve impulse is one of most actively
investigated problems in the history of Neuroscience \cite{Hodgk1,Hodgk2,nag,Hodgk3,Hodgk4,Tas,Hodgk5a,Hodgk5,Hodgka,Hodgk6a,Hodgk6a1,Hodgk6a2,Hodgk6a3,Hodgk6a4,Hodgk6,bonalex}. Study of the problem is motived by the crucial need for a good understanding of characteristic properties of the action potential, assumed to be a propagating form of transmembrane voltage along the axon. Measurements of action potentials in several physiological contexts have generated a wealth of data that triggered a great deal of theoretical interest in the phenomenon. As pioneer in this theoretical interest, the Hodgkin-Huxley model \cite{Hodgk1,Hodgk2} rests on a picture by which the nerve impulse is an electric voltage propagating in form of an asymmetric pulse
along the nerve fiber. Originally the Hodgkin-Huxley model was introduced to explain data obtained from measurements of conductive parameters of a nerve fiber, and particularly to show how these data could be used to directly calculate both the shape and velocity of an action potential on the squid giant axon \cite{cole}. \par
According to the Hodgkin-Huxley model \cite{Hodgk1}, the nerve impulse is a
self-regenerative wave associated with the electrochemical
activity of the nerve cell, and due to the flow of ion currents (Na$^+$
and K$^+$) through specific ion channels. This wave propagates with a
constant shape, through a mechanism that can be summarized as follow: During the generation and
transmission of the nerve impulse, the leading edge of the
depolarization region of the nerve triggers adjacent membranes to depolarize,
causing a self-propagation of the excitation related to the
transmembrane voltage down the nerve fiber
\cite{Hodgk1,warman,grill}. Hodgkin and Huxley suggested that a
convenient way to describe the propagation of this transmembrane voltage is to regard the nerve fiber as an electric cable. Thus,
in its most conventional formulation, the Hodgkin-Huxley model assumes currents in intracellular and extracellular fluids to be ohmic such that the net transmembrane current is the sum of ionic and capacitive currents. In this picture the conservation law for currents passing through the membrane can be written \cite{Hodgk1}:
\begin{equation}\label{e1}
    C_m\frac{\partial V}{\partial t}=D\frac{\partial^2
    V}{\partial x^2}-F(V),
\end{equation}
where $V$ is the transmembrane voltage, $C_m$ is the membrane capacitance, $D$ is the diffusion coefficient and $F$ accounts for contributions from some ion currents.
\par 
Still, besides the indisputable electrical activity of the axonal membrane, experiments have also pointed out \cite{mueller,hem,Tasaki2,Tasaki1,blunk1} the existence of mechanical constraints related to pressures due to flows of fluids through the membrane. Concretely these mechanical constraints are electromechanical forces that are responsible for mechanotransduction processes \cite{gross}, physiological processes in which mechanical forces such as pressures exerted by ionic fluids on cell membranes and tissues, can trigger excitations of electrical natures playing important role in the control of various stimuli-responsive organs, in homeostasis of living organisms and so on \cite{gross}. 
\par Taking advantage of experiments suggesting sizable thermodynamic phenomena preceeding and following the action potential, and specifically the liquid-gel transition observed at some critical temperature \cite{nat1,nat2,nat3,nat4}, Heimburg and Jackson \cite{Hodgk5a,theim,heimburgs} suggested that mechanical forces related to pressure waves could play a major role in nerve membrane excitation and subsequently in the buildup of nerve impulse. In this respect they postulated that pressure waves associated with propagation of the density difference between fluids flowing through the nerve membrane, could actually be a mechanical manifestation of the action potential.
Most recently there have been few other attempts to revisit the mathematical description of the nerve impulse, with a main aim to combine the contributions of electrical and mechanical processes \cite{gross,nat1,nat2,nat4,eng}.\par
In this work we propose a model describing the electromechanical process of generation of the action potential. The model combines the Hodgkin-Huxley cable model and the pressure-wave model proposed by Heimburg and Jackson \cite{Hodgk5a}. Our model assumes that the membrane capacitance changes instantanously with the difference in densities of fluids through the membrane, leading to a modified Hodgkin-Huxley equation where the membrane capacitor now behaves like a "feedback" component (i.e. like a capacitive diode). \par
In sec. \ref{sec:two} we present the model which consists of two nonlinear partial differential equations, namely the modified Hodgkin-Huxley equation for the action potential and the Boussinesq equation for the density-difference wave \cite{Hodgk5a,theim}. In sec. \ref{sec:three} we first consider the stationary regime of the action-potential (or modified Hodgkin-Huxley) equation. In this purpose we use the exact soliton solution to the Korteweg-de Vries (KdV) equation derived from Boussinesq's equation, to recast the modified Hodgkin-Huxley equation into a linear operator problem with zero eigenvalue. Three exact bound-state solutions to this linear operator problem are obtained analytically, for specific values of characteristic parameters of the model. In sec. \ref{sec:four}, numerical simulations of the modified Hodgkin-Huxley equation are carried out assuming the three stationary solutions as initial profiles of the action potential. In sec. \ref{sec:five} we conclude the study.
\section{\label{sec:two} The model}
\smallskip
The axon can be regarded as a long cylinder with walls made of cell
membrane surrounded by intracellular and extracellular fluids \cite{Hodgk2,eng}. The
intracellular fluid stands for a conductive liquid with a high
concentration of potassium ions but a low concentration of sodium
and chlorine ions, while the axonal cell membrane acts like a barrier
preventing ions in the intracellular liquid from mixing with
external solutions. Due to the difference in ion concentrations in intracellular and extracellular fluids, a resting potential is expected to set up through the membrane. If the nerve is
depolarized, e.g. due to the presence of a stimulus of any kind, the axon membrane will become selectively permeable to ionic currents which flow rapidly into the cell, reversing the polarity of the action potential \cite{Hodgk1,Hodgk2}.\par
In general, for a fixed number of charged lipids around the cell membrane, the charge density will be different because the respective
lipid areas are different \cite{gross}. Therefore we can expect changes in the electrostatic potential of the membrane during a propagating
pressure wave, indicating a possibility of an
electromechanical coupling between the net fluid density and
the electrostatic potential on the cell membrane. This electromechanical coupling, first reported by
Petrov \cite{Petrov} and widely observed in recent experiments in neurophysiology \cite{gross,nat1,nat2,grossa,grossb,grossc}, can also be linked with changes in membrane capacitance as a result of variation of the fluid density through the membrane. \par
The model proposed in this study retains the key ingredients \cite{Hodgk1} of the Hodgkin-Huxley cable model, except for the self-regulatory function of the membrane capacitance now assumed to vary instantaneously with the net fluid density on the membrane. With this consideration, the system dynamics can be described by the following set of two nonlinear space-time partial differential equations:
\begin{eqnarray}\label{e5}
D\frac{\partial^2 V}{\partial x^2}&=& \frac{\partial}{\partial t} \bigg( C_m(x,t) \,V\bigg), \label{e5a}\\
\frac{\partial^2 U}{\partial
    t^2}&=&c_0^2\frac{\partial}{\partial
    x}\Bigg((1-U)\frac{\partial U}{\partial
    x}\Bigg)-h\frac{\partial^4 U}{\partial x^4}. \label{e5b}
\end{eqnarray}
In eq. (\ref{e5b}) we introduced a dimensionless variable $U=\Delta\rho^A/\rho_0$ to represent the density difference $\Delta\rho^A=\rho^A-\rho_0^A$, note that physical meanings of parameters $\Delta\rho^A$, $\rho_0^A$ and $\rho_0$ are discussed in detail in refs. \cite{Hodgk5a,hem,heimburgs,theim}. \par To describe the instantaneous change of the membrane capacitance $C_m$ due to variation of the ion-carrying fluid density \cite{Tasaki2,Tasaki1}, we postulate that when the nerve is active the rate of change of the membrane capacitance is proportional to the net density of ion-carrying fluid $\Delta\rho^A$ on the membrane i.e.:
 \begin{equation}\label{e6a}
 \frac{\partial C_m(x,t)}{\partial t}=\kappa \Delta\rho^A,
 \end{equation}
where $\kappa$ is assumed positive. Using eq. (\ref{e6a}) the modified Hodgkin-Huxley equation (\ref{e5a}) becomes:
\begin{equation}\label{e6c}
C_m(x,t)\frac{\partial V}{\partial t}=D\frac{\partial^2 V}{\partial
x^2}-\kappa\Delta\rho^A(x,t)V, \end{equation}
where the membrane capacitance $C_m(x,t)$ is given by:
\begin{equation}\label{e6d}
C_m(x,t)=C_0+\kappa \int{\Delta\rho^A(x,t)dt}.
\end{equation}
Instructively the value $\kappa=0$ reproduces the standard Hodgkin-Huxley model \cite{Hodgk1,Hodgk2}, however for nonzero values of $\kappa$ eq. (\ref{e6c}) turns to a modified Hodgkin-Huxley equation whose solution depends on the spatio-temporal profile of the density-difference wave $\Delta\rho^A(x,t)$. In the next section, using the exact one-soliton solution to eq. (\ref{e5b}), we seek for possible analytical solutions to the modified Hodgkin-Huxley equation (\ref{e5a}). In this respect we shall see that the modified Hodgkin-Huxley equation is analytically tractable only in the steady-state regime. Indeed in this regime the modified Hodgkin-Huxley equation reduces to a linear-operator problem with zero eigenvalue, the bound states of which are Legendre polynomials \cite{abram}. 
\section{\label{sec:three} The action potential in stationary regime}
\smallskip
By introducing new coordinates;
\begin{equation}
U(x,t)=\psi(\xi,T), \hskip 0.25truecm \xi=\frac{c}{c_0}(x-c_0\,t), \hskip 0.25truecm
T=\frac{h}{c}t,
\end{equation}
and integrating once with respect to the new variable $\xi$, eq. (\ref{e5b}) reduces to the KdV equation \cite{ggkm}:
\begin{equation}\label{e8}
\frac{\partial \psi}{\partial T}=\alpha\psi\frac{\partial
\psi}{\partial\xi}-\beta\frac{\partial^3 \psi}{\partial \xi^3},
\end{equation}
where $\alpha$ and $\beta$ are constants depending
on $c_0$, $h$ and $c$. The parameters $\alpha$ and $\beta$ can be set to any values through judicious coordinate transformations, however we shall retain the most widely used values of these parameters namely $\alpha=6$ and $\beta=-1$ \cite{ggkm}. For these specific values the KdV eq. (\ref{e8}) admits exact one and n-soliton solutions, which are obtained by means of the inverse-scattering transform \cite{ggkm}. Focusing on the one-soliton solution, the inverse-scattering transform suggests the following analytical expression: 
 \begin{equation}\label{e15d}
     U(x,t)=-2sech^2(x-4t),
 \end{equation}
 which is a localized wave of depression. \par
\begin{figure*} 
\begin{minipage}[c]{0.32\textwidth}
\includegraphics[width=1.7in,height=1.7in]{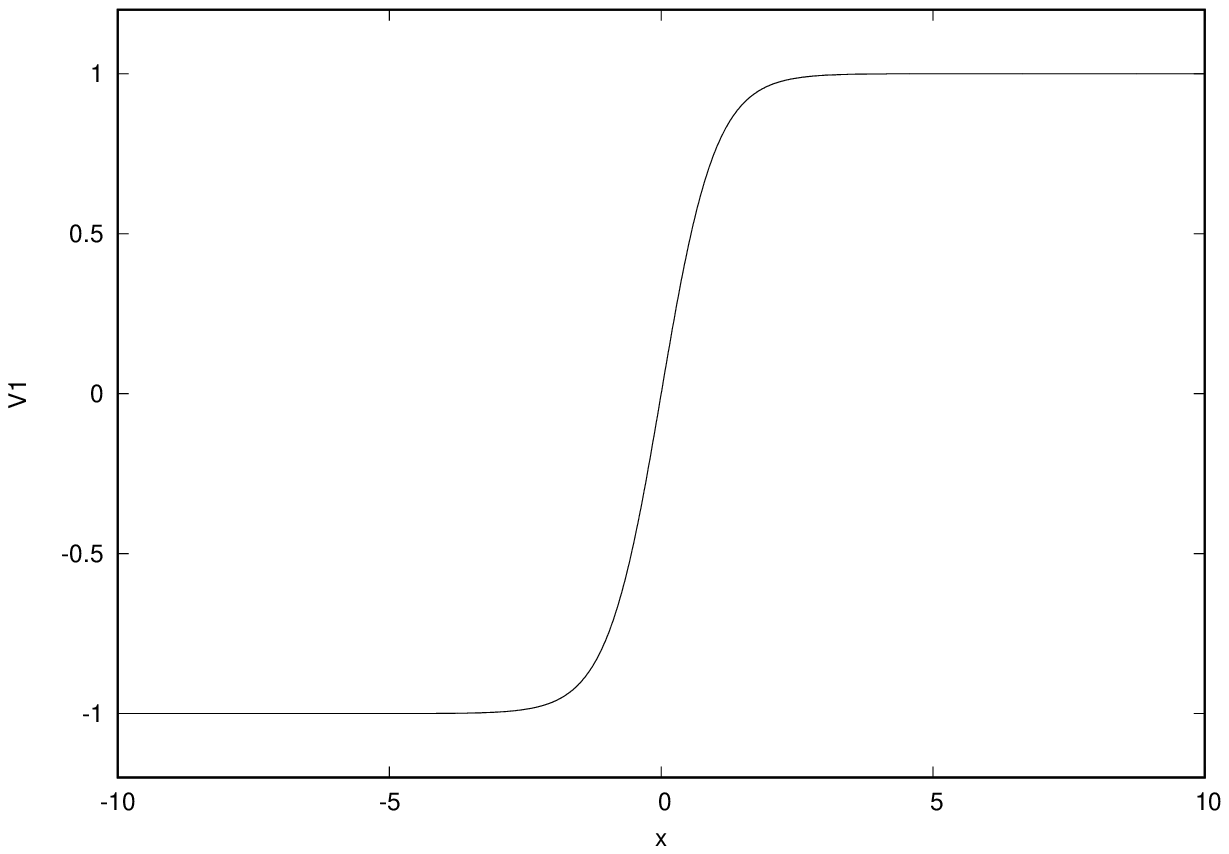}
\end{minipage}%
\begin{minipage}[c]{0.32\textwidth}
\includegraphics[width=1.7in,height=1.7in]{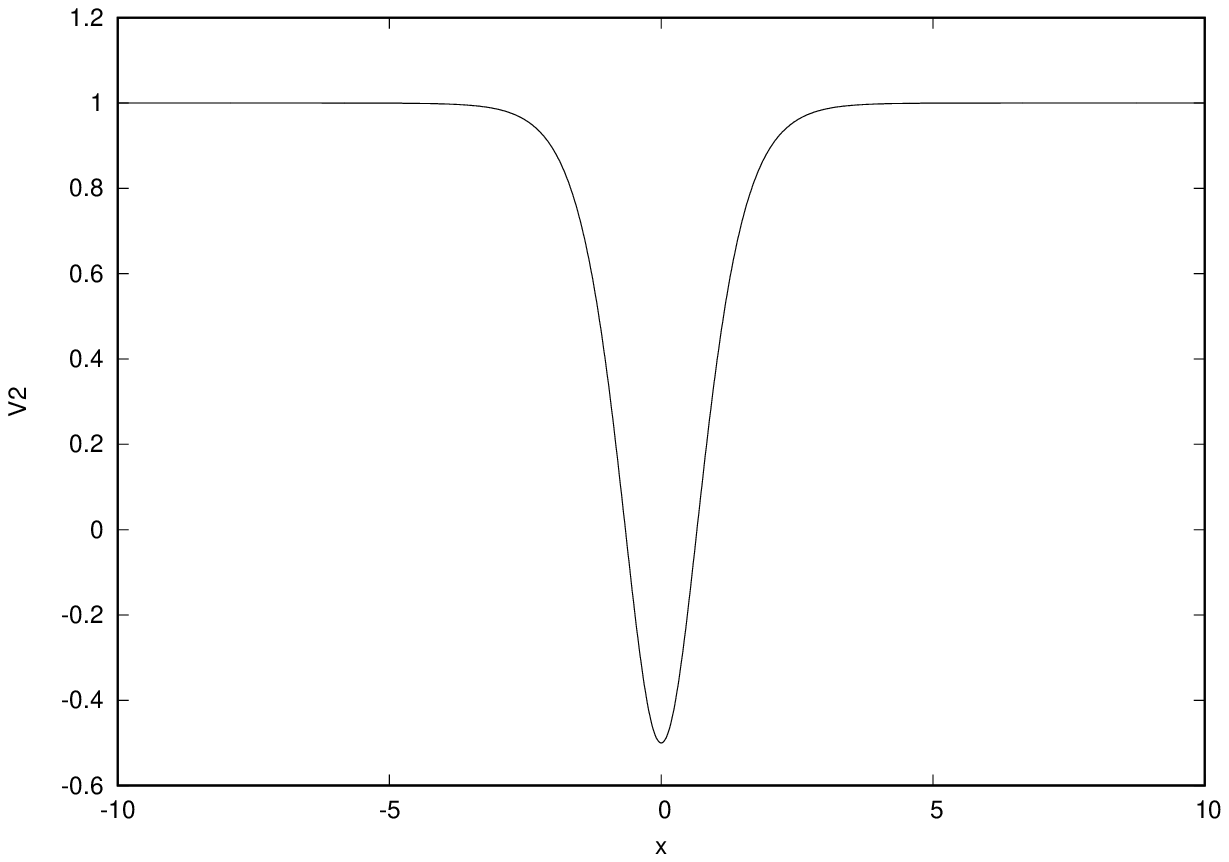}
\end{minipage}%
\begin{minipage}[c]{0.32\textwidth}
\includegraphics[width=1.7in,height=1.7in]{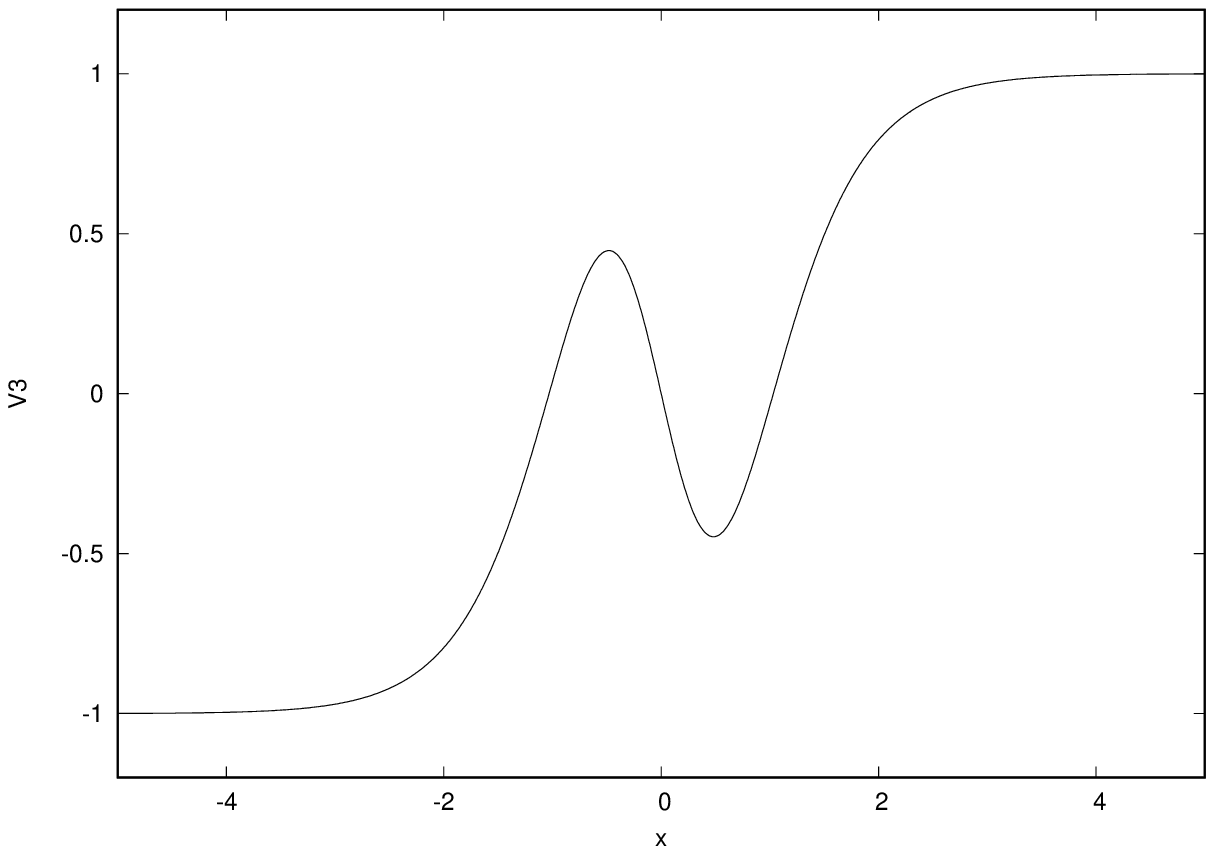}
\end{minipage} 
\caption{\label{fig:one} Sketches of the first three bounded modes of the zero-eigenvalue equation (\ref{e13}). From left to right: $n=1$, $2$, $3$.} 
\end{figure*}

 With the help of the one-pulse solution (\ref{e15d}) we can re-express the modified Hodgkin-Huxley equation (\ref{e6c}) as:
 \begin{equation}\label{e11}
C_m(x,t)\frac{\partial V}{\partial t}=D\frac{\partial^2 V}{\partial
x^2}-\varepsilon U(x,t)V, 
\end{equation}
with $\varepsilon=\kappa\rho_0$. Equation (\ref{e11}) needs to be fully solved in order to gain a consistent picture of the spatio-temporal evolution of the action potential $V(x,t)$. Unfortunately this equation is complex as it stands, and no exact solution can be obtained except via numerical simulations. Neverthless we remark that at steady state, this equation reduces to an eigenvalue problem for which exact analytical solutions can be found for specific values of $\mu$. To this last point, in steady-state regime eq. (\ref{e11}) reduces to the zero-eigenvalue linear operator problem:        
\begin{equation}\label{e13}
\Bigg( \vartheta(x) -\frac{\partial^2}{\partial x^2} \Bigg) V(x)=0, \hskip 0.3truecm \vartheta(x)= -\mu\, sech^2 x,
\end{equation}
in which $\mu=\frac{2\varepsilon}{D}$, where we used the simplified notation $V(x)=V(x,0)$. By defining $\tau=\tanh x$, eq. (\ref{e13}) can be transformed into a Legendre equation of order n i.e. \cite{abram}:
\begin{equation}
    \frac{d}{d\tau}\left\{(1-\tau^2)\frac{dV}{d\tau}\right\}+n(n+)V=0, \label{jacob}
 \end{equation}
where $n(n+1)=\mu$, $n$ being a positive integer. Bounded solutions to the linear operator equation (\ref{e13}), for an arbitrary $n$, are the Legendre polynomials: 
\begin{equation}\label{e15e}
V_n(\tau)=\frac{1}{2^n n!}\frac{d^n}{d\tau^n}(\tau^2-1)^n, \hskip 0.25truecm n=1,2,3, \cdots.
\end{equation}
For illustration, below we list the three lowest bounded modes: 
\begin{eqnarray}
V_1(x)&=&\tanh x, \hskip 0.5truecm  D=2\epsilon, \label{bste1} \\
V_2(x)&=&\frac{1}{2}(3 \tanh^2 x - 1), \hskip 0.5truecm  D=6\epsilon, \label{bste2} \\
V_3(x)&=&\frac{1}{2}(5 \tanh^2 x -3)\tanh x, \hskip 0.5truecm  D=12\epsilon, \label{bste3}
\end{eqnarray}
which are sketched in fig. (\ref{fig:one}).
In the next section we shall proceed to numerical simulations of the modified Hodgkin-Huxley equation (\ref{e11}), using the above three bounded modes as input profiles $V(x, 0)$ of the action potential (i.e. as initial conditions).
\section{\label{sec:four} Numerical solutions to the variable-capacitance Hodgkin-Huxley equation}
\smallskip
The modified Hodgkin-Huxley equation (\ref{e11}) is an initial-value problem, as such it can be solved numerically using a finite-difference algorithm. In our case we adopt a finite-difference scheme that combines a central-difference approximation for the time derivative and a forward-difference approximation for the second-order derivative in space \cite{land}. As we are interested more in a qualitative analysis than a quantitative description of the problem, values of the diffusion coefficient $D$, the bare membrane capacitance $C_0$, the electromechanical coupling coefficient $\kappa$ and the quantity $\epsilon$ will be arbitrary. \par
\begin{figure}\centering
 \includegraphics[width=3.2in,height=4.1in]{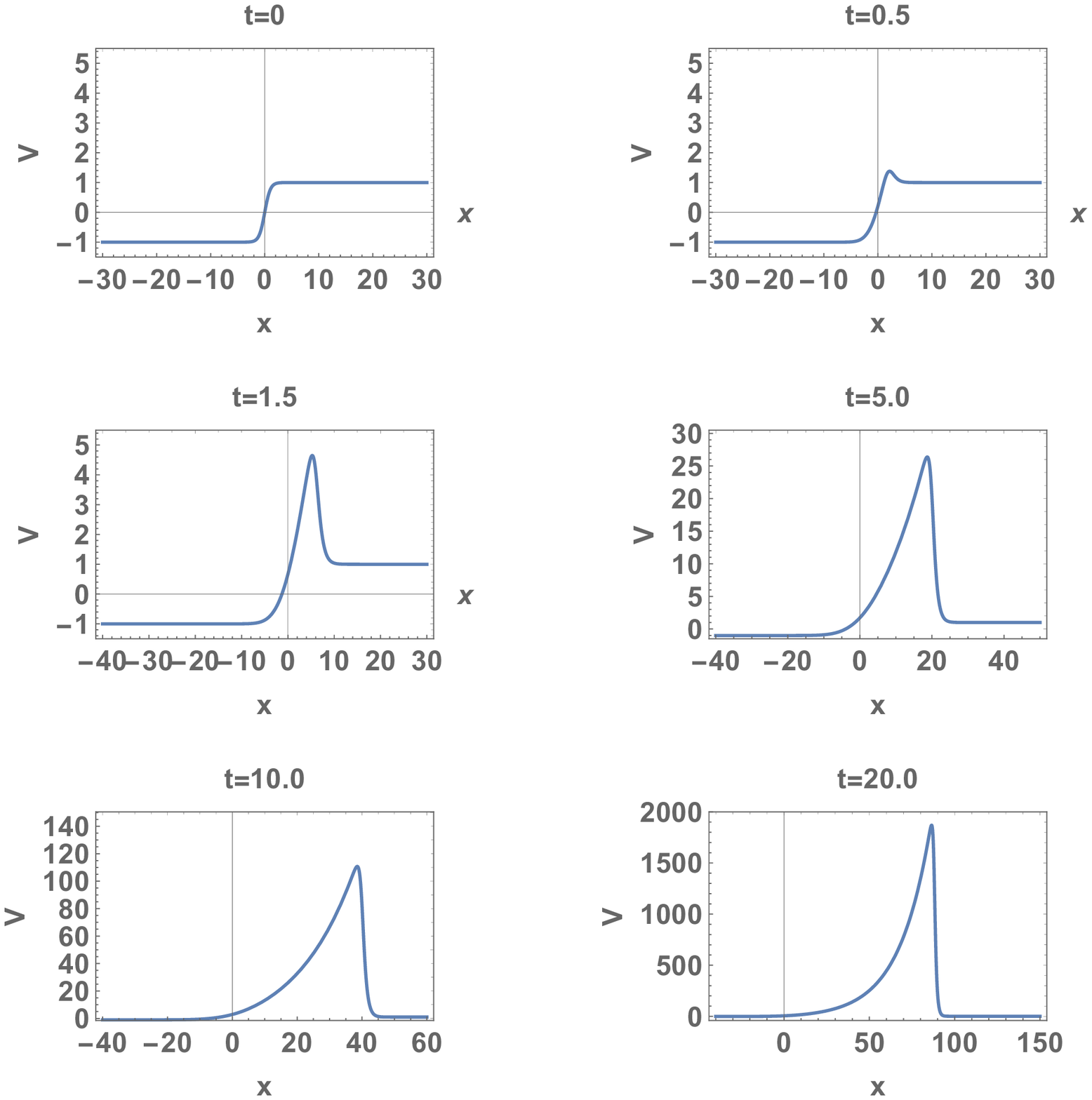}
 \caption{\label{figa}(Color online) Profiles of the transmembrane voltage at different times $t$, obtained from numerical simulations of eq. (\ref{e11}) with the bound state eq. (\ref{bste1}) used as initial solution: $D=6.0$, $C_0=2.0$, $\kappa=0.3$, $\epsilon =3.3$.} 
 \end{figure}
\begin{figure}\centering 
 \includegraphics[width=3.2in,height=4.1in]{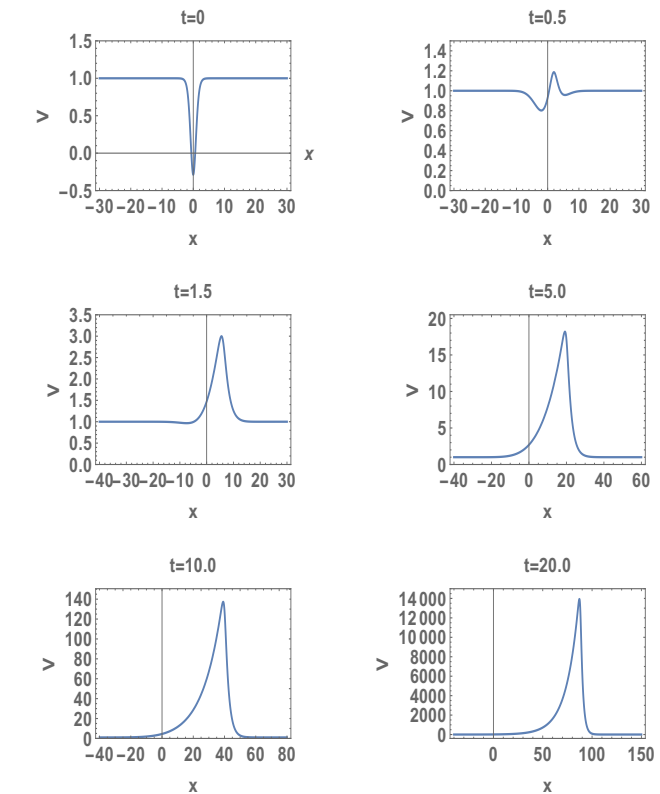}
 \caption{\label{figb}(Color online) Profiles of the transmembrane voltage at different times $t$, obtained from numerical simulations of eq. (\ref{e11}) with the bound state eq. (\ref{bste2}) used as initial condition: $D=6.0$, $C_0=2.0$, $\kappa=0.3$, $\epsilon =3.3$.} 
 \end{figure}
\begin{figure}\centering 
 \includegraphics[width=3.2in,height=4.1in]{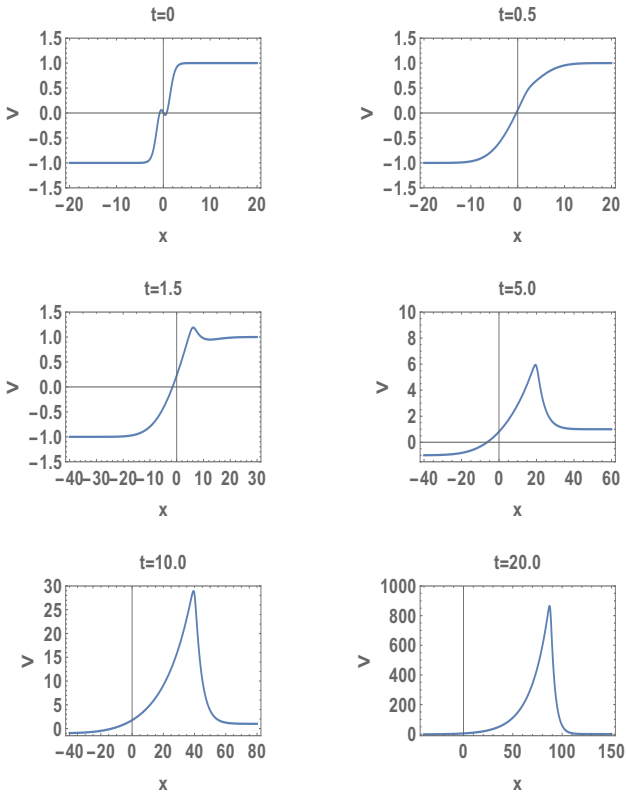}
 \caption{\label{figc}(Color online) Profiles of the transmembrane voltage at different times $t$, obtained from numerical simulations of eq. (\ref{e11}) with the bound state eq. (\ref{bste3}) used as initial condition: $D=6.0$, $C_0=2.0$, $\kappa=0.3$, $\epsilon =3.3$.} 
 \end{figure}

Graphs in figs. \ref{figa}, \ref{figb} and \ref{figc} represent profiles of the transmembrane voltage $V(x,t)$ at six different times $t$, generated numerically from the modified Hodgkin-Huxley equation (\ref{e11}) for the three distinct initial conditions (\ref{bste1}) (fig. \ref{figa}), (\ref{bste2}) (fig. \ref{figb}) and (\ref{bste3}) (fig. \ref{figc}). Values of parameters are $D=6$, $C_0=2$, $\kappa=0.3$ and $\epsilon=3.3$.
\par
\begin{figure}\centering 
 \includegraphics[width=3.in,height=3.1in]{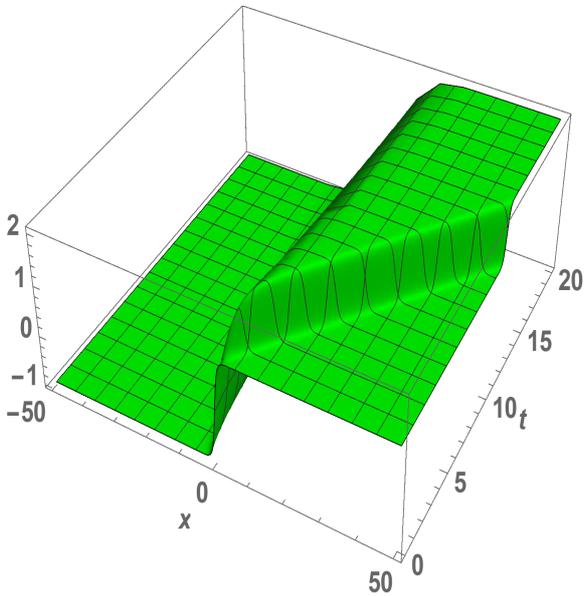}
 \caption{\label{figg}(Color online) Spatio-temporal shape of the transmembrane voltage $V(x,t)$, obtained numerically with the initial profile eq. (\ref{bste1}): $\epsilon=2$, $\kappa=0.3$, $C_0=2.9$.} 
 \end{figure}

\begin{figure}\centering 
 \includegraphics[width=3.in,height=3.1in]{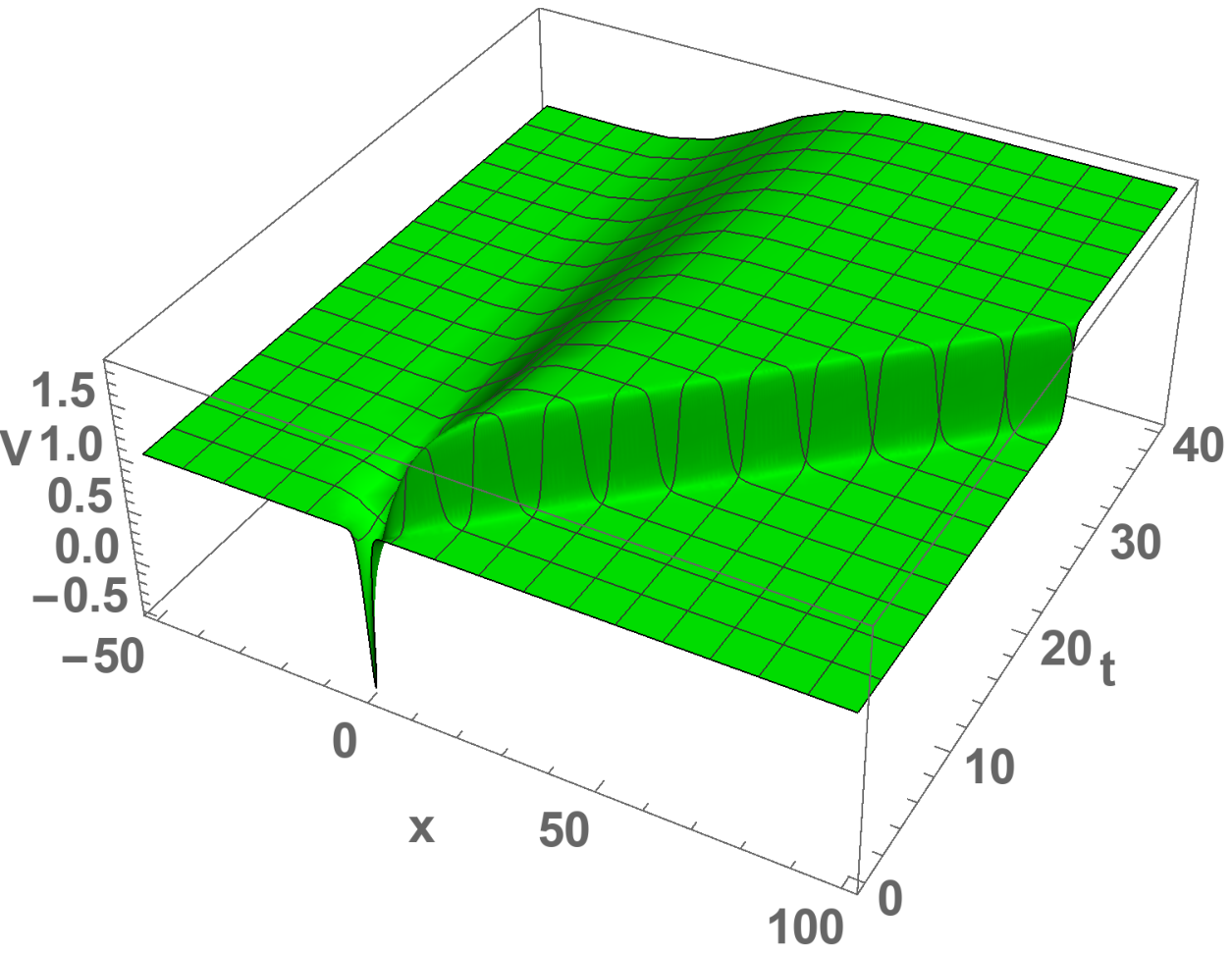}
 \caption{\label{figh}(Color online) Spatio-temporal shape of the transmembrane voltage $V(x,t)$, obtained numerically with the initial profile eq. (\ref{bste2}): $\epsilon=2$, $\kappa=0.3$, $C_0=2.9$.} 
 \end{figure}

\begin{figure}\centering 
 \includegraphics[width=3.in,height=3.1in]{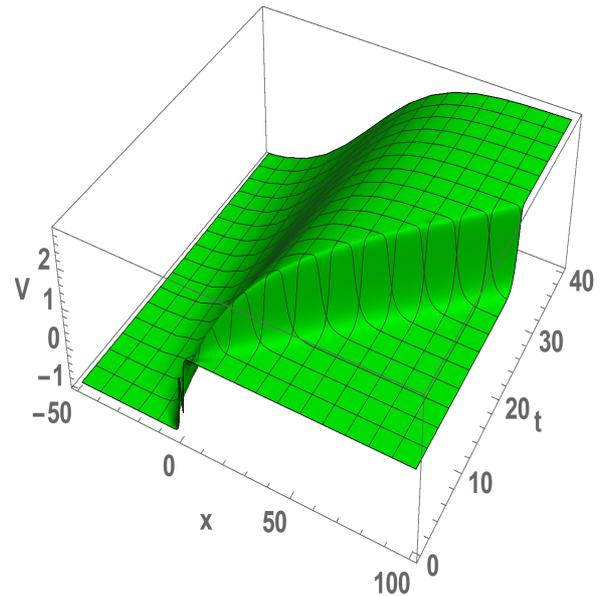}
 \caption{\label{figi}(Color online) Spatio-temporal shape of the transmembrane voltage $V(x,t)$, obtained numerically with the initial profile eq. (\ref{bste3}): $\epsilon=2$, $\kappa=0.3$, $C_0=2.9$.} 
 \end{figure}

 In fig. \ref{figa}, which is the numerical solution with the kink bound state (\ref{bste1}) as initial condition, it is seen that after some propagation time, the kink-shaped input is modulated and stabilizes permanently in a typical pulse shape characteristic of the action potential \cite{breta,ya,yab}.
 Graphs of figs. \ref{figb} and \ref{figc} exhibit the same feature as fig. \ref{figa}, meaning that the two other initial solutions will also stabilize in a pulse shape similar to fig. \ref{figa} after a transient propagation time.
To gain a global view of the spatio-temporal evolution of the three different solutions shown in the previous figures, they were represented in three dimensions as depicted in figs. \ref{figg}, \ref{figh} and \ref{figi}. The three-dimensional representation clearly indicate that as they propagate along the axon, the three different initial profiles always modulate into the same pulse pattern.
\section{\label{sec:five} Conclusion}
\smallskip
The mechanism by which the nerve impulse is generated and transmitted along the axon has been a fundamental problem not only in neurophysiology, but also in mathematical physics\cite{Hodgk6a1,Hodgk6a2,Hodgk6a3,Hodgk6a4}. In their pioneer model, Hodgkin and Huxley \cite{Hodgk1} suggested a picture based on an electric cable according to which the nerve impulse would be a propagating form of a depolarization, due to ion exchanges through the axon membrane. However in view of the undeniable role of membrane in selectively passing ion from intracellular to extracellular fluids and vice-versa, Hodgkin-Huxley's electric-cable picture was subsequently improved by regarding the membrane capacitance as a feedback organ \cite{Hodgk5}. Observations of thermodynamic phenomena in the nerve activity, such as the heat release during liquid-gel transition with subsequent generation of acoustic waves along the axon \cite{Hodgk5a,theim}, motivated a distinct picture involving mechanical processes related to a variable density difference of liquids flowing through the nerve membrane. This new picture \cite{Hodgk5a} led to the idea that pressure waves could be a manifestation of the action potential.\par
For the electrical \cite{Hodgk1} and mechanical \cite{Hodgk5a} pictures, taken separately, enable only partial descriptions of the process of nerve impulse generation, the need for a description taking simultaneously into consideration the electrical and mechanical activities of the cell membrane, was an imperative. In this study we exploited experimental evidences of electromechanical phenomena, and their specific manifestations in some neurophysioical contexts \cite{gross}, to introduce a model which rests on the Hodgkin-Huxley electrical model, but assumes the membrane capacitance to be determined by the difference in densities of ion-carrying fluids flowing accross the membrane. The proposed model combines the KdV equation already present in the soliton model of Heimburg and Jackson \cite{Hodgk5a}, and the Hodgkin-Huxley's electric cable equation with a feedback capacitor (i.e. a capacitive diode). By postulating a mathematical expression describing the relationship between the membrane capacitance and the density difference, we found that in steady-state regime the action potential equation reduces to a zero-eigenvalue linear operator problem. This linear operator problem can be transformed into the Legendre equation, the solutions of which is the family of Legendre polynomials. The three lowest bound states of this linear operator equation were obtained, and used as initial profiles in numerical simulations of the full partial differential equation describing the spatio-temporal evolution of the transmembrane voltage. From numerical simulations it turned out that the three distinct initial profiles always decay into the a common pulse profi;e after a transient propagation time. \par
To end, let us underline that the model proposed in the present study, can be improved to account several relevant aspects of the process that we neglected. For instance in the modified Hodgkin-Huxley equation (\ref{e5a}), we ingored the contribution of ion currents $F(V)$ yet this term plays an important role the original Hodgkin-Huxley model \cite{Hodgk1,Hodgk2}. Also it is known that the Boussinesq equation describing the dynamics of pressure waves, is actually obtained from a Tyalor expansion of the fluid velocity field with respect to the density difference $\Delta \rho^A$ (or $U$). It is well established that carrying out the expansion beyond the linear term, leads to higher-order or modified KdV equations \cite{teem} which admit soliton solutions distinct from eq. (\ref{e15d}). Therefore including these quantities in the present model will undoubtedly enrich qualitatively the physics of the process under study. In particular having distinct soltion solutions for the density wave implies distinct bound-state spectra, and hence new profiles for the action potential that might possibly be more close to the reality.

\section*{Conflict of interest}
The author declares that he has no conflict of interest.
%

\end{document}